\documentclass[twocolumn,preprintnumbers,showpacs,amsmath,amssymb]{revtex4}
\usepackage{graphicx}
\usepackage{dcolumn}
\usepackage{bm}
\begin{document}
\title{Multiple-transit paths and density correlation functions in PASEP}
\author{Farhad H. Jafarpour}
\email{farhad@ipm.ir}
\author{Somayeh Zeraati}
\affiliation{Physics Department, Bu-Ali Sina University, 65174-4161 Hamedan, Iran}
\date{\today}
\begin{abstract}
We consider the partially asymmetric simple exclusion process (PASEP) when its steady-state probability distribution function can be written in terms of a linear superposition of product measures with a finite number of shocks. In this case the PASEP can be mapped into an equilibrium walk model, defined on a diagonally rotated square lattice, in which each path of the walk model has several transits with the horizontal axis. We particularly show that the multiple-point density correlation function in the PASEP is related to the probability that a path has multiple contacts with the horizontal axis from the above or below.
\end{abstract}
\maketitle
\section{Introduction}
The Partially Asymmetric Simple Exclusion Process, or PASEP for short, usually refers to a system of identical classical particles moving on a discrete lattice. In this system the particles are injected and extracted from both ends of the lattice while hopping to the left and right in the bulk of the lattice with unequal rates $x$ and $1$ respectively. The particles perform continuous time random walks on the lattice. The exclusion rule forbids two particles to occupy a single lattice site at the same time \cite{Ligget1,Ligget2}.\\
The steady-state of the PASEP has been widely studied in the related literatures. The exact results have been obtained using a matrix product approach for the PASEP with open boundaries \cite{TS-open,BECE} and also on a ring in the presence of a second class particle which is sometimes called a defect particle \cite{TS-ring}; however, not much is known about the dynamics of shocks which are defined as discontinuities in the density profile of the particles in the PASEP. In \cite{BS02,KJS03,JM07} the authors have studied the time evolution of a product measure, as an initial probability distribution, with a finite number of shocks. It has been shown that at any time $t \ge 0$ the probability distribution function can be written as a linear combination of shock measures with an structure similar to that of the initial configuration. On the other hand, the shock positions perform continuous time random walks interacting by the exclusion rule. The Monte-Carlo simulations for a single shock confirm these results \cite{BCFG89,KSKS98}.\\
It can be simply understood that if the dynamics of the shock position in a one-dimensional driven-diffusive system is similar to that of a simple random walker which reflects from the boundaries of the system, then the steady-state of the system can be written in terms of a linear superposition of these product measures. On the other hand, it has been shown that the steady-state of such system can be written in terms of a matrix element of a product of non-commuting operators which satisfy an algebra. According to this approach, known as the matrix product method, one associates an operator to each state of a lattice site. The steady-state of the system in this case is written in terms of a matrix element (for the open boundaries conditions) or a trace (for the periodic boundary conditions) of a product of such non-commuting operators which might have finite- or infinite-dimensional matrix representations. A review of this method can be found in \cite{BE07}.\\
It is known that if the steady-state of a one-dimensional driven-diffusive system with open boundaries can be written in terms of a linear superposition of product measures containing a single shock, then the steady-state of the system can be equivalently expressed by using the matrix product method with two-dimensional matrix representations. It has also been found that the shock characteristics appear in the matrix elements of the matrix representations i.e. the matrix elements are functions of the shock hoping rates and the densities of the particles on the left and the right hand sides of the shock position \cite{JM07}. As the authors in \cite{JM07} have proposed that this idea can be easily generalized to the PASEP, where an $N$-dimensional matrix representation for the quadratic algebra of this system exists, which is surprisingly valid under the same constraints under which the steady-state of the system can be written as a superposition of product measures with $N-1$ shocks \cite{BS02}. The main difference between this matrix representation with those proposed before is that it is clearly written in terms of the hoping rates of the shock fronts and also the densities of the particles separated by the them.\\
The steady-state of the PASEP has been recently studied from different viewpoints. In \cite{BCEP06} a combinatorial derivation and interpretation of the weights, associated with the stationary distribution of the PASEP, has been introduced. By using this approach the authors have found explicit expressions for the stationary distribution and normalization using both recurrences and path models. In a recent work the authors in \cite{BJJK09} have used a continued fraction representation of the lattice path generating function to discuss the steady-state properties of the PASEP. In this case an infinite-dimensional tridiagonal matrix representation of the quadratic algebra of the PASEP has been used to calculate the grand-canonical normalization of the stationary distribution of the system. It has been shown that by truncating these continued fractions one can recover the results obtained from a finite-dimensional matrix representations of the quadratic algebra which are valid only along special lines in the phase diagram of this system \cite{KS97}.\\
Apart from these approaches one can try to connect the matrix representation of the algebra of a one-dimensional reaction-diffusive system directly to the transfer matrix of an equilibrium two-dimensional walk model. In this way the one can actually show that the normalization coefficient of the stationary distribution function of the non-equilibrium system coincides with the partition function of an equilibrium walk model \cite{BE04,BGR04,JZ10}.\\
In this paper we consider the PASEP with open boundaries where the steady-state of the system can be written in terms of a linear superposition of product measures with a finite number of shocks. This means that the quadratic algebra of the system has a finite-dimensional representation under some constraints on the reaction rates. It is known that the $N$-dimensional representations of the stationary algebra describe the stationary linear combination of shock measures with $N-1$ consecutive shocks \cite{KJS03}. Considering an $N$-dimensional matrix representation of the quadratic algebra we show that the partition function of the PASEP is equal to that of a two-dimensional walk model on a diagonally rotated square lattice. The partition function of the walk model is the sum of the weighted paths which start and end on the horizontal axis and cross it $N-1$ times. The multiple-point density correlation functions of the PASEP can be written in terms of the equilibrium characteristics of the  walk model. By connecting the finite-dimensional matrix representation of the PASEP to the transfer matrix of the weighted walk model, we can easily calculate the multiple-point density correlation function of the PASEP in terms of physical quantities defined for the equilibrium walk model.\\
In order to write a self-explanatory paper we have organized it as follows: first we will start with the definition of the PASEP with open boundaries and explain how the stationary distribution of this system can be written using a matrix product method. We will then introduce the multiple-transit walk model and find its partition function using a transfer matrix method. The multiple-point density correlation functions of the PASEP are explicitly calculated at the end of the manuscript.
\section{The PASEP}
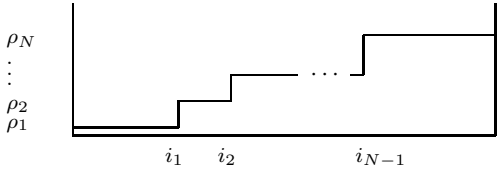
\begin{figure}
\begin{center}
\begin{picture}(150,70)
\put(10,10){\line(0,1){50}}
\put(10,10){\line(1,0){160}}
\put(170,10){\line(0,1){50}}
\put(10,13){\line(1,0){40}}
\put(50,13){\line(0,1){10}}
\put(50,23){\line(1,0){20}}
\put(70,23){\line(0,1){10}}
\put(70,33){\line(1,0){25}}
\put(115,33){\line(1,0){5}}
\put(100,31){$\cdots$}
\put(120,33){\line(0,1){15}}
\put(120,48){\line(1,0){50}}
\put(45,0){\footnotesize $i_{1}$}
\put(65,0){\footnotesize $i_{2}$}
\put(117,0){\footnotesize $i_{N-1}$}
\put(-15,13){\footnotesize $\rho_{1}$}
\put(-15,20){\footnotesize $\rho_{2}$}
\put(-15,29){\footnotesize $\vdots$}
\put(-15,45){\footnotesize $\rho_{N}$}
\end{picture}
\caption{Sketch of the particle distribution containing $N-1$ shocks.}
\label{fig1}
\end{center}
\end{figure}

We consider a discrete lattice of length $L$. Each lattice site can be occupied by one particle or is empty. Occupying a single lattice site with two particles is forbidden. In the bulk of the system each particle hops to the left (right) neighboring site with the rate $x$ ($1$) provided that it is not already occupied. The particles can enter into the system from the leftmost (rightmost) lattice site with the rate $(1-x)\alpha$ ($(1-x)\delta$). The particles can also leave the system from the leftmost (rightmost) lattice site with the rate $(1-x)\gamma$ ($(1-x)\beta$). The above-mentioned process, called the PASEP, has been studied widely in related literatures.\\
Defining a shock as a sharp discontinuity in the density profile of the particles on the lattice, it has been shown that a product shock measure with $N-1$ consecutive shocks defined as follows
\begin{widetext}
\begin{equation}
\label{PSM}
\vert i_{1},i_{2},\cdots,i_{N-1} \rangle=\left( \begin{array}{c} 1-\rho_1 \\
\rho_1 \end{array} \right)^{\otimes i_{1}} \otimes \left( \begin{array}{c} 1-\rho_2 \\
\rho_2 \end{array} \right)^{\otimes i_{2}-i_{1}}\otimes \cdots \otimes \left( \begin{array}{c} 1-\rho_N \\
\rho_N \end{array} \right)^{\otimes i_{N-1}-L}
\end{equation}
\end{widetext}
can evolve in time according to ($N-1$)-particle dynamics provided that the reaction rates lie on a specific surface in the parameters space defined by $\alpha,\;\beta,\;\gamma,\;\delta$ and $x$. The dynamics of each shock position in this case will be similar to that of a simple random walker moving on a discrete lattice which reflects from the boundaries with special rates. In Fig. \ref{fig1} we have sketched an initial distribution of particles on the lattice with $N-1$ shocks locating at different lattice sites.
We have assumed that the $i$'th shock hops to the left (right) with the rate $\delta_{l,i}$ ($\delta_{r,i}$) for $i=1,\cdots,N-1$. In this case the densities and the shock hopping rates satisfy the following relations \cite{BS02,KJS03}
\begin{equation}
x=\frac{\rho_{i}(1-\rho_{i+1})}{\rho_{i+1}(1-\rho_{i})} \;\; \mbox{for} \;\; i=1,\cdots,N-1
\end{equation}
and also
\begin{equation}
\frac{\delta_{r,i}}{\delta_{l,i}}=\frac{\rho_{i+1}(1-\rho_{i+1})}{\rho_{i}(1-\rho_{i})} \;\; \mbox{for} \;\; i=1,\cdots,N-1.
\end{equation}
Using the matrix product method the stationary probability distribution vector $\vert P^*\rangle$ for the PASEP on a lattice of length $L$ can be written as
\begin{equation}
\label{MPSS}
\vert P^* \rangle =\frac{1}{Z} \langle\langle W \vert \left( \begin{array}{c} D_{0} \\
D_{1} \end{array} \right)^{\otimes L} \vert V \rangle\rangle
\end{equation}
in which $Z$ is a normalization factor and can be obtained easily using the normalization condition
\begin{equation}
\label{NOR}
Z_{L,N}=\langle\langle W \vert C^L\vert V \rangle\rangle
\end{equation}
in which we have defined $C:=D_{0}+D_{1}$. The operator $D_{0}$ stands for the presence of a vacancy and the operator $D_{1}$ stands for the presence of a particle at each lattice site. These operators besides the vectors $\langle\langle W \vert$ and $\vert V \rangle\rangle$ satisfy the following quadratic algebra
\begin{equation}
\label{PASEPalg}
\begin{array}{l}
D_{1}D_{0}-x  D_{0}D_{1}=\xi(1-x)(D_{0}+D_{1})\\
\langle\langle W \vert (\alpha D_{0} - \gamma D_{1}- \xi)=0\\
(\beta D_{1}- \delta D_{0}-\xi)\vert V \rangle\rangle=0
\end{array}
\end{equation}
in which $\xi=\rho_{N}(1-\rho_{N})$. It can be easily verified that the following $N$-dimensional matrices
\begin{widetext}
\begin{eqnarray}
D_{0}=\left( \begin{array}{cccccc}
\frac{\rho_{N}(1-\rho_{N})}{\rho_1} &\frac{\rho_{N}(1-\rho_{N})}{\rho_2} & 0& \cdots& 0 &0\\
0&\frac{\rho_{N}(1-\rho_{N})}{\rho_2}&\frac{\rho_{N}(1-\rho_{N})}{\rho_3}&\cdots&0&0\\
0&0&\frac{\rho_{N}(1-\rho_{N})}{\rho_3}& \cdots&0&0\\
\vdots &\vdots&\vdots& \ddots&\vdots&\vdots\\
0&0&0&\cdots&0&\frac{\rho_{N}(1-\rho_{N})}{\rho_{N}}
\end{array} \right),
D_{1}=\left( \begin{array}{cccccc}
\frac{\rho_{N}(1-\rho_{N})}{1-\rho_1} &\frac{\rho_{N}(1-\rho_{N})}{1-\rho_2} & 0& \cdots& 0 &0\\
0&\frac{\rho_{N}(1-\rho_{N})}{1-\rho_2}&\frac{\rho_{N}(1-\rho_{N})}{1-\rho_3}&\cdots&0&0\\
0&0&\frac{\rho_{N}(1-\rho_{N})}{1-\rho_3}& \cdots&0&0\\
\vdots &\vdots&\vdots& \ddots&\vdots&\vdots\\
0&0&0&\cdots&0&\frac{\rho_{N}(1-\rho_{N})}{1-\rho_{N}}
\end{array} \right)
\end{eqnarray}
\end{widetext}
besides the vectors
$$
\begin{array}{ll}
\vert V \rangle\rangle = \left( \begin{array}{c}
v_1\\
v_2\\
\vdots\\
v_N
\end{array} \right) \;\; \mbox{and}\;\;
\langle\langle W \vert = \left( \begin{array}{cccc}
w_1&w_2&\cdots&w_N
\end{array} \right)
\end{array}
$$
satisfy (\ref{PASEPalg}) provided that we have
\begin{equation}
\begin{array}{l}
w_N=1\;,\;w_i=\frac{\prod_{j=i+1}^{N}(\alpha-(1+\alpha+\gamma-\rho_j)\rho_j)}
{\prod_{j=i+1}^{N}(-\alpha+(\alpha+\gamma)\rho_j)}\;\mbox{for}\;i\neq N
\end{array}
\end{equation}
and
\begin{equation}
\begin{array}{l}
v_1=1\;,\;v_i=\frac{\prod_{j=1}^{i-1}(1-\frac{\beta}{1-\rho_j}+\frac{\delta}{\rho_j})}
{\prod_{j=2}^{i}(\frac{\beta}{1-\rho_j}-\frac{\delta}{\rho_j})}\;\mbox{for}\;i\neq 1.
\end{array}
\end{equation}
The boundary rates should also satisfy the following constraint
\begin{equation}
x^{N-1}=\kappa_{+}(\alpha,\gamma)\kappa_{+}(\beta,\delta)
\end{equation}
in which we have defined
\begin{equation}
\kappa_{+}(u,v)=\frac{-u+v+1+\sqrt{(u-v-1)^2+4uv}}{2u}.
\end{equation}
Using the boundary conditions one finds
\begin{equation}
\rho_1=\frac{1}{1+\kappa_{+}(\alpha,\gamma)}\;\;,\;\;
\rho_N=\frac{\kappa_{+}(\beta,\delta)}{1+\kappa_{+}(\beta,\delta)}.
\end{equation}
It is interesting to study the properties of the matrix $C$ in this representation which is given by
\begin{equation}
\label{C}
C=\left( \begin{array}{cccccc}
\frac{\rho_N(1-\rho_N)}{\rho_1(1-\rho_1)}&
\frac{\rho_N(1-\rho_N)}{\rho_2(1-\rho_2)}&0&\cdots&0&0\\
0&\frac{\rho_N(1-\rho_N)}{\rho_2(1-\rho_2)}&\frac{\rho_N(1-\rho_N)}{\rho_3(1-\rho_3)}&\cdots&0&0\\
0&0&\frac{\rho_N(1-\rho_N)}{\rho_3(1-\rho_3)}&\cdots&0&0\\
\vdots &\vdots&\vdots& \ddots&\vdots&\vdots\\
0&0&0&\cdots&0&1
\end{array} \right).
\end{equation}
It can be seen the $i$th diagonal element of the matrix $C$ is equal to $\frac{\rho_N(1-\rho_N)}{\rho_i(1-\rho_i)}=\prod_{j=i}^{N-1}\frac{\delta_{r,j}}{\delta_{l,j}}$. Considering the case $0 \le \rho_1<\rho_2<\cdots<\rho_N \le 1$, the reader can convince himself that only two different cases (phases) exist in the steady-state. In the first phase we have $\frac{\rho_N(1-\rho_N)}{\rho_1(1-\rho_1)}>1$ and in the second phase we have $\frac{\rho_N(1-\rho_N)}{\rho_1(1-\rho_1)}<1$. On the coexistence line one has $\frac{\rho_N(1-\rho_N)}{\rho_1(1-\rho_1)}=1$. It is known that if one starts with the product shock measure (\ref{PSM}), the shock positions move with constant speed until the two shocks meet and then coalesce into a single shock. In a long-time limit only one shock (the leftmost, which is the fastest) survives \cite{FFV00}. The particle densities on the left- and the right-hand sides of this shock are equal to $\rho_1$ and $\rho_N$ respectively. The quantity $\frac{\rho_N(1-\rho_N)}{\rho_1(1-\rho_1)}$ gives the ratio of the hopping rate of the shock position to the right and left. In the first (second) phase the shock position has more of a tendency to move to the right (left), and therefore the bulk density will be equal to $\rho_1$ ($\rho_N$).\\
In the next section we will introduce an equilibrium multiple-transit walk model and show how it can be connected to the PASEP in the above-mentioned case.
\section{The multiple-transit walk model}
\begin{figure}
\begin{center}
\begin{picture}(150,70)
\put(-30,50){\line(1,0){200}}
\put(-30,50){\line(1,1){10}}
\put(-20,60){\line(1,-1){10}}
\put(-10,50){\line(1,1){10}}
\put(0,60){\line(1,-1){10}}
\put(10,50){\line(1,-1){10}}
\put(20,40){\line(1,1){10}}
\put(30,50){\line(1,1){10}}
\put(40,60){\line(1,-1){10}}
\put(50,50){\line(1,1){10}}
\put(60,60){\line(1,-1){10}}
\put(70,50){\line(1,1){10}}
\put(80,60){\line(1,-1){10}}
\put(90,50){\line(1,-1){10}}
\put(100,40){\line(1,1){10}}
\put(110,50){\line(1,-1){10}}
\put(120,40){\line(1,1){10}}
\put(130,50){\line(1,1){10}}
\put(140,60){\line(1,-1){10}}
\put(150,50){\line(1,1){10}}
\put(160,60){\line(1,-1){10}}
\put(-40,40){\footnotesize $(0,0)$}
\put(160,40){\footnotesize $(2L,0)$}
\put(-15,65){\footnotesize $1st$}
\put(15,30){\footnotesize $2nd$}
\put(55,65){\footnotesize $3rd$}
\put(105,30){\footnotesize $4th$}
\put(145,65){\footnotesize $5th$}
\end{picture}
\caption[fig2]{A typical multiple-transit path containing $5$ Dyck paths and $4$ transits.}
\label{fig2}
\end{center}
\end{figure}
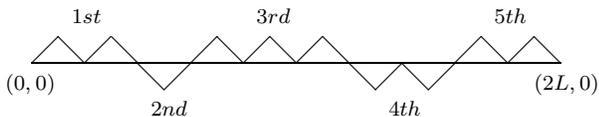

We consider a two-dimensional walk model on a diagonally-rotated square lattice. A random walker moves along a path which starts at ($0,0$) and ends at ($2L,0$) (see Fig. \ref{fig2}). The random walker starts from $(0,0)$ and is allowed to move one step in the North-East (NE) or in the South-East (SE) direction. It can move at most two consecutive steps to the NE or to the SE only when it crosses the horizontal axis. All other movements are prohibited. We study the case where the random walker can cross the horizontal axis at most $N-1$ times. In this case the paths taken by the random walker might cross the horizontal axis $m$ times where $m=0,\cdots,N-1$; therefore, the path contains $m+1$ Dyck paths. We label each Dyck path with $l$ where $l=1,\cdots,N$. The $l$th Dyck path lies above (below) the horizontal axis if $l$ is odd (even).
We assign an statistical weight to each path (which might contain different combinations of Dyck paths with different lengths) taken by the random walker. The partition function of this multiple-transit walk model is then the sum of these statistical weights. The statistical weight of a path is equal to the multiplication of the weights of the individual Dyck paths times the weight of the first step times the weight of the last step. In order to find the statistical weight of a path of length $2L$ with $m$ transits, the following steps should be taken. The statistical weight associated with the $l$th Dyck path is defined as follows: if $l$ is an even number, we assign the fugacity $\frac{1}{z_{l}}$ to each contact with the horizontal axis except to the last upward step, and if $l$ is an odd number, we assign the fugacity $\frac{1}{z_{l}}$ to each contact with the horizontal axis, except the first upward step. If the path starts with the $1$st Dyck path, we assign the fugacity $\frac{1}{\widetilde{z}_1}$ to the contact of the first step with the horizontal axis (equivalently the point ($0,0$)); otherwise, if the path starts with the $l$th Dyck path ($l \neq 1$), two different scenarios might happen: it is either the first step of the $l$th Dyck path (which means that it comes from a transit) or the second step and above. For the first case we assign a fugacity $\frac{1}{\widetilde{z}_{l-1}}$ to the contact of the first step with the horizontal axis (equivalently the point ($0,0$)) or $\frac{1}{\widetilde{z}_{l}}$ for the latter case. We always assign a fugacity $\frac{1}{\widetilde{z}'_{l}}$ to the contact of the last step with the horizontal axis if the path ends to the $l$th Dyck path.\\
Defining the weight of a path as described above one finds (after some straightforward but lengthly calculations)
the partition function of the walk model as follows:
\begin{equation}
\label{WMPF}
\widetilde{Z}_{L,N}=\sum_{l=1}^{N}[\frac{z_{l}^{-L}}{\widetilde{z}_{l}\widetilde{z}'_{l}}+\sum_{j=1}^{l-1}
\sum_{k=j}^{l}[\frac{z_{j}z_{k}^{-L+l-j-1}}{\widetilde{z}_{j}\widetilde{z}'_{l}\prod_{i=j,i\neq k}^{l}(z_i-z_k)}]].
\end{equation}
Up to this point we have not considered any constraints on the fugacities. First of all one can easily see that the fugacities associated with the boundaries i.e. $\widetilde{z}_i$'s and $\widetilde{z}'_i$'s do not play any major role in the thermodynamic limit ($L \rightarrow \infty$) of the partition function (\ref{WMPF}); however, one can consider the cases in which either $z_1$ or $z_N$ is the smallest bulk fugacity and all other fugacities lie between them. This results in two different behaviors for the partition function of the walk model in the thermodynamic limit. In the first (second) case where $z_1$ ($z_N$) is the smallest bulk fugacity, the majority of paths contain the Dyck path of the first (last) type $l=1$ ($l=N$). This defines two different phases which will be discussed later. In what follows we will show how the partition function (\ref{WMPF}) can be obtained using the transfer matrix method.
\section{The transfer matrix method}
In this section we will show that the partition function of the walk model (\ref{WMPF}) can be calculated using a transfer matrix formalism. By defining a two-step transfer matrix $T=T^oT^e$, in which $T^o$ and $T^e$ are associated with odd and even steps taken by the the random walker respectively, we will show that the partition function of the model (\ref{WMPF}) can be written as
\begin{equation}
\label{TM}
\widetilde{Z}_{L,N}=\langle L \vert T^L \vert R \rangle.
\end{equation}
One can easily find the transfer matrix $T$ and the vectors $\langle L \vert$ and $\vert R \rangle$ as follows: we first define a complete base vector associated with the vertices of every path.
\begin{figure}
\begin{center}
\begin{picture}(140,100)
\put(0,50){\line(1,0){60}}
\put(70,50){\line(1,0){60}}
\put(0,50){\line(1,1){30}}
\put(30,80){\line(1,-1){30}}
\put(70,50){\line(1,-1){30}}
\put(100,20){\line(1,1){30}}
\put(-2,60){\footnotesize $\downarrow$}
\put(-10,70){\footnotesize $\vert 0^{2l-1} \rangle$}
\put(28,85){\footnotesize $\downarrow$}
\put(25,95){\footnotesize $\vert 1^{2l-1} \rangle$}
\put(58,60){\footnotesize $\downarrow$}
\put(50,70){\footnotesize $\vert 0^{2l-1} \rangle$}
\put(68,40){\footnotesize $\uparrow$}
\put(60,30){\footnotesize $\vert 0^{2l} \rangle$}
\put(98,10){\footnotesize $\uparrow$}
\put(95,0){\footnotesize $\vert 1^{2l} \rangle$}
\put(128,40){\footnotesize $\uparrow$}
\put(120,30){\footnotesize $\vert 0^{2l} \rangle$}
\end{picture}
\caption[fig3]{Assigning the base vectors to the vertices.}
\label{fig3}
\end{center}
\end{figure}
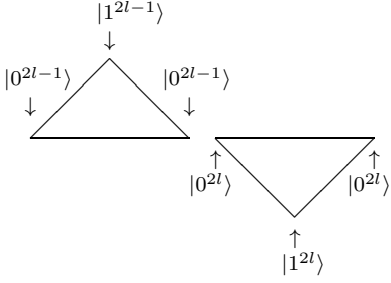
The base vector $\vert 0^{l} \rangle$ is assigned to a contact point with zero height. For the $l$th Dyck path, the contact point will be from the below (above) the horizontal axis, if $l$ is even (odd). On the other hand, the base vector $\vert 1^{l} \rangle$ is assigned to a point with a unit height either above or below the horizontal axis. For the $l$th Dyck path if $l$ is even (odd), then the point lies below (above) the horizontal axis. An example is given in Fig. \ref{fig3}. These base vectors are complete in the following sense
\begin{equation}
\sum_{l=1}^{N} \vert 1^{l} \rangle \langle 1^{l} \vert=
\sum_{l=1}^{N} \vert 0^{l} \rangle \langle 0^{l} \vert=\cal{I}
\end{equation}
in which $\cal{I}$ is an $N\times N$ identity matrix. For instance one can choose the following matrix representation for the base vectors
\begin{equation}
\vert 1^{l} \rangle=
\vert 0^{l} \rangle=
  \left(\begin{array}{c}
  0 \\ \vdots \\ 0 \\ 1\\ 0 \\ \vdots \\ 0
  \end{array}\right)_{N \times 1} \;\;\mbox{for}\;\;l=1,\cdots,N
\end{equation}
in which only the $l$th element is non-zero and equal to one. If we write the partition function (\ref{WMPF}) as
\begin{widetext}
\begin{equation}
\widetilde{Z}_{L,N}=\sum_{l_1=1}^{N}\cdots\sum_{l_{L+1}=1}^N\langle L \vert 0^{l_1}\rangle\langle 0^{l_1}\vert T^{o}\vert 1^{l_2}\rangle\langle 1^{l_2}\vert T^{e} \vert 0^{l_3} \rangle\langle 0^{l_3}\vert \cdots\vert 0^{l_{L+1}}\rangle\langle 0^{l_{L+1}} \vert R \rangle
\end{equation}
\end{widetext}
then according to the definition of our walk model the only non-zero transitions are given by
\begin{equation}
\label{ME}
\begin{array}{l}
\langle 0^{l} \vert T^{o} \vert 1^{l'} \rangle =\delta_{l,l'}+\delta_{l,l'-1},\\
\langle 1^{l} \vert T^{e} \vert 0^{l'} \rangle =\frac{1}{z_l}\delta_{l,l'}\\
\langle L\vert 0^{l} \rangle=\frac{1}{\widetilde{z}_l},\\
\langle 0^{l} \vert R \rangle=\frac{1}{\widetilde{z}'_l}
\end{array}
\end{equation}
in which $\delta_{i,j}$ is the usual Kronecker delta. These relations imply that we should have the following matrices
\begin{equation}
\begin{array}{ll}
\label{MR1}
T^o=\left(
\begin{array}{ccccc}
1 & 1 & \cdots & 0 & 0\\
0 & 1 & \cdots & 0 & 0\\
\vdots & \vdots  & \ddots & \vdots & \vdots\\
0 & 0  & \cdots & 1 & 1\\
0 & 0 &  \cdots & 0 & 1
\end{array} \right),&
T^e=\left(
\begin{array}{ccccc}
\frac{1}{z_{1}} & 0  & \cdots & 0 & 0\\
0 & \frac{1}{z_{2}}  & \cdots & 0 & 0\\
\vdots & \vdots  & \ddots & \vdots & \vdots\\
0 & 0 & \cdots & \frac{1}{z_{N-1}} & 0\\
0 & 0 & \cdots & 0 & \frac{1}{z_{N}}
\end{array} \right)
\end{array}
\end{equation}
and vectors
\begin{equation}
\begin{array}{ll}
\label{MR2}
\langle L \vert = (\frac{1}{\widetilde{z}_1},\cdots,\frac{1}{\widetilde{z}_N}) \; , &
\vert R \rangle= \left( \begin{array}{c}
\frac{1}{\widetilde{z}'_1} \\ \vdots\\ \frac{1}{\widetilde{z}'_N} \end{array} \right).
\end{array}
\end{equation}
It is now easy to verify that using these matrix representation, besides the transfer matrix representation of the partition function (\ref{TM}), one obtains (\ref{WMPF}).\\
An interesting physical quantity which can be defined is the probability that the $l$th Dyck path has a contact at site $2i$ with the horizontal axis given by
\begin{equation}
\label{OPCF}
\langle P_{i}^{l}\rangle_{L}=\frac{\langle L \vert T^{i} \widehat{P}_{i}^{l}T^{L-i}\vert R \rangle}{\langle L \vert T^{L}\vert R \rangle},
\end{equation}
in which the contact operator $\widehat{P}_{i}^{l}$ is defined as follows
\begin{equation}
\widehat{P}_{i}^{l}=\vert 1^{l}\rangle_{2i-1}\langle 0^{l}\vert_{2i}.
\end{equation}
Normalization requires us to have
\begin{equation}
\sum_{l=1}^{N}\langle P_{i}^{l}\rangle_{L}=1.
\end{equation}
One can easily generalize this idea to find the probability that $m$ Dyck paths $l_1, l_2, \cdots, l_m$ ($l_1<\cdots<l_N$) have contacts with the horizontal axis at sites $2i_1, 2i_2, \cdots, 2i_m$ ($i_1<\cdots<i_m$). The associated contact operator in this case can be written as
\begin{widetext}
\begin{equation}
\widehat{P}_{i_1,i_2,\cdots,i_m}^{l_1,l_2,\cdots,l_m}=\vert 1^{l_1}\rangle_{2i_1-1}\langle0^{l_1}\vert_{2i_1}T^{i_2-i_1}\vert 1^{l_2}\rangle_{2i_2-1}\langle0^{l_2}\vert_{2i_2}T^{i_3-i_2}\vert1^{l_3}\rangle_{2i_3-1} \cdots \langle0^{l_{m-1}}\vert_{2i_{m-1}}T^{i_m-i_{m-1}}\vert 1^{l_m}\rangle_{2i_m-1}\langle0^{l_m}\vert_{2i_m}.
\end{equation}
\end{widetext}
Now the appropriate probability is given by
\begin{equation}
\label{MPCF}
\langle P_{i_1,\cdots,i_m}^{l_1,\cdots,l_m}\rangle_{L}=\frac{\langle L \vert T^{i_1} \widehat{P}_{i_1,\cdots,i_m}^{l_1,\cdots,l_m}T^{L-i_m}\vert R \rangle}{\langle L \vert T^{L}\vert R \rangle}
\end{equation}
in which
\begin{equation}
\sum_{l_1,l_2,\cdots,l_m=1}^{N}\langle P_{i_1,\cdots,i_m}^{l_1,\cdots,l_m}\rangle_{L}=1.
\end{equation}
We have assumed that either $z_1$ or $z_N$ is the smallest bulk fugacity of the walk model. The thermodynamic behavior of (\ref{MPCF}) will be then interesting under these conditions. After some calculation one finds in large $L$ limit
\begin{widetext}
\begin{equation}
\label{Contact1}
\langle P_{i_1,\cdots,i_m}^{l_1,\cdots,l_m}\rangle_{L}\cong
\left\{
\begin{array}{ll}
( 1-\{\frac{\frac{\widetilde{z}'_1}{\widetilde{z}'_N}}{\frac{z_N}{z_1}+\frac{\widetilde{z}'_1}{\widetilde{z}'_N}
-1} \}(\frac{z_N}{z_1})^{i_m-L})\prod_{k=1}^{m} \delta_{l_k,1} \;\; \mbox{for} \; \frac{z_1}{z_N}<1,\\ \\
(1+\{\frac{\frac{\widetilde{z}_N}{\widetilde{z}_1}}{\frac{z_N}{z_1}-\frac{\widetilde{z}_N}{\widetilde{z}_1}
-1} \}(\frac{z_N}{z_1})^{i_1}) \prod_{k=1}^{m} \delta_{l_k,N} \;\; \mbox{for} \; \frac{z_1}{z_N}>1.
\end{array}
\right.
\end{equation}
\end{widetext}
It can be seen that these probabilities have exponential behaviors with a correlation length which is determined by $z_1$ and $z_N$. As explained above, for $\frac{z_1}{z_N}<1$ ($\frac{z_1}{z_N}>1$) it is more probable that the paths contain the first $l=1$ (last $l=N$) Dyck path. On the coexistence line where $z_1=z_N$ one finds that the paths might contain both the first $l=1$ and the last $l=N$ Dyck paths of any length not larger than $2L$. On this line and in the large $L$ limit the probability (\ref{MPCF}) is given by the following expression which is obviously linear in terms of the position of contacts
\begin{equation}
\label{Contact2}
\langle P_{i_1,\cdots,i_m}^{l_1,\cdots,l_m}\rangle_{L}\cong\frac{1}{L}(i_{k}-i_{k-1})\prod_{{i'}=1}^{k-1}\delta_{l_{i'},1}\prod_{{i''}=k}^{m}\delta_{l_{i''},N}
\end{equation}
for $k=1,\cdots,m+1$ assuming that $i_0=0$ and $i_{m+1}=L$. This expression indicates that on the coexistence line all the probabilities are zero unless the paths only contain two types of Dyck paths $l=1$ and $l=N$ and that the first $k-1$ contacts belong to the first Dyck path and the last $m-k+1$ contacts belong to the last Dyck path.\\
In the next section we will show how the multiple-transit random walk model is related to the PASEP and that how the multiple-point density correlation functions can be written in terms of the physical quantities of the walk model (\ref{OPCF}) and (\ref{MPCF}).
\section{Density correlation functions in the PASEP}
The reader can easily see that the transfer matrix $T$ in (\ref{TM}) has the following $N\times N$ matrix representation
\begin{equation}
T=T^oT^e=\left(
\begin{array}{ccccc}
\frac{1}{z_{1}} & \frac{1}{z_{2}}  & \cdots & 0 & 0\\
0 & \frac{1}{z_{2}}  & \cdots & 0 & 0\\
\vdots & \vdots  & \ddots & \vdots & \vdots\\
0 & 0 & \cdots & \frac{1}{z_{N-1}} & \frac{1}{z_{N}}\\
0 & 0 & \cdots & 0 & \frac{1}{z_{N}}
\end{array} \right).
\end{equation}
Comparing this with (\ref{C}) one can recognize that they are the same matrices provided that
\begin{equation}
\frac{1}{z_i}=\frac{\rho_{N}(1-\rho_N)}{\rho_{i}(1-\rho_i)}
\end{equation}
for $i=1,\cdots,N-1$ and that $z_N=1$. On the other hand the partition function of the PASEP (\ref{NOR}) will be equal to that of the multiple-transit walk model (\ref{WMPF}) provided that $v_i=\frac{1}{\widetilde{z}'_i}$ and $w_i=\frac{1}{\widetilde{z}_i}$ for $i=1,\cdots,N$. In this case one has $C=T$, $\vert V \rangle \rangle=\vert R \rangle$ and $\langle \langle W \vert =\langle L \vert$.\\
This mapping allows us to write the physical quantities of the two systems. For instance the mean density of the particles at a given site $i$ in the PASEP $\langle \rho_i \rangle_L$, written in terms of the matrix product formalism as
\begin{equation}
\langle \rho_i \rangle_L=\frac{\langle\langle W \vert C^{i-1}D_1C^{L-i} \vert V \rangle\rangle}{\langle\langle W \vert C^L\vert V \rangle\rangle}
\end{equation}
can be rewritten as
\begin{equation}
\langle \rho_i \rangle_L=\sum_{l=1}^{N}\rho_l \langle P_{i}^{l} \rangle_{L}
\end{equation}
by noticing that $D_1=\sum_{l=1}^{N}\rho_l T \widehat{P}_{i}^{l}$. Generalizing this to the $m$-point density correlation function for the PASEP one finds
\begin{equation}
\label{MPDCF}
\langle \rho_{i_1}\cdots \rho_{i_m} \rangle_L=\sum_{l_1,\cdots,l_m=1}^{N}\rho_{l_1}\cdots\rho_{l_m} \langle P_{i_1,i_2,\cdots,i_m}^{l_1,l_2,\cdots,l_m}\rangle_{L}.
\end{equation}
Considering the thermodynamic behavior of (\ref{MPCF}) given by (\ref{Contact1}) and (\ref{Contact2}) one can calculate the thermodynamic behavior of the multiple-point density correlation function (\ref{MPDCF}) for the PASEP. In the same way one can also calculate the connected multiple-point density correlation functions. It can be seen that for the case of two point correlation functions the results of \cite{ER96} can be recovered. For instance on the coexistence line one finds
\begin{equation}
\langle \rho_{Lx_1}\cdots \rho_{Lx_m} \rangle_L=\sum_{k=1}^{m+1}(x_{k}-x_{k-1}){\rho_1}^{k-1}{\rho_N}^{m-k+1}
\end{equation}
where we have defined $x_k=\frac{i_k}{L}$ and this is exactly the same expression as calculated in \cite{MS97}.
\section{Concluding remarks}
Recently, connections between some of the one-dimensional driven-diffusive systems with the equilibrium walk models have been under intensive investigations. In this direction it is interesting to show how the physical quantities in equilibrium systems are related to those in non-equilibrium systems.\\
In this paper we have shown that the multiple-point density correlation functions in the PASEP are explicitly related to the probabilities of having multiple contacts with the horizontal axis of a path in an equilibrium two-dimensional walk model. In order to show this relation we have adopted a matrix product approach while using an special finite-dimensional matrix representation for the quadratic algebra of the system which explicitly reveals the shock characteristics of the PASEP. This matrix representation helps us connect the partition function of the PASEP to that of a multiple-transit walk model obtained using the transfer matrix method.\\
There are a couple of examples in related literature which show that the stationary distribution of the asymmetric simple exclusion process (ASEP) with open boundaries can be related to that of some path models. It has been shown that the partition function of each path model, obtained using the transfer matrix method, is equal (or at least proportional) to that of the ASEP, calculated using the matrix product method.\\
As we have seen in this paper, some of the physical quantities defined in both models are related to each other using these methods. However, it is still not clear why and how these methods are related and that whether ASEP is an exception or one can relate different one-dimensional driven-diffusive systems to the path models using these methods. On the other hand, one can also consider the driven-diffusive systems with periodic boundary conditions and investigate whether this connection still holds.

\end{document}